\documentclass[12pt]{article}
\usepackage{amssymb,amsmath,epsfig}
\begin{document}
\title{\bf Gravitational Dust Collapse in $f(R)$ Gravity}

\author{M. Farasat
Shamir$^{(1)}$\thanks{farasat.shamir@nu.edu.pk}, Zahid Ahmad
$^{(2)}$\thanks{zahidahmad@ciit.net.pk} and Zahid Raza$^{(1)}$ \thanks{zahid.raza@nu.edu.pk} \\\\  $^{(1)}$Department of
Sciences \& Humanities, \\National University of Computer \&
Emerging Sciences,\\ Lahore Campus, Pakistan.\\ $^{(2)}$Department of Mathematics, COMSATS,\\Institute of Information Technology,
University Road,\\ Abbottabad, Pakistan.}

\date{}

\maketitle
\begin{abstract}
This paper is devoted to investigate gravitational collapse of dust
in metric $f(R)$ gravity. We take FRW metric for the
interior region while the Schwarzchild spacetime is considered for
the exterior region of a star. The junction
conditions have been derived to match interior and exterior
spacetimes. The assumption of constant scalar curvature
is used to find a solution of field equations. Gravitational mass is found
by using the junction conditions. It is concluded that the constant
curvature term $f(R_0)$ plays the role of the cosmological constant
involved in the field equations of general relativity.
\end{abstract}


\section{Introduction}

The most interesting topic in the gravitational physics today is
the expansion of our universe. The support to this argument comes from different sources
such as Supernovae Ia (SNIa) experiments \cite{sn}, cosmic microwave
background fluctuations \cite{2} and X-ray experiments \cite{xray}. All these
observations indicate that our universe is expanding with an
accelerated rate. The phenomenon of dark matter and dark energy
is another topic of discussion \cite{de1}.
Einstein gave the concept of dark energy in $1917$ by
introducing a small positive cosmological constant in the field equations. But later on,
he rejected it. However, it is now believed that the cosmological constant may be a suitable
candidate for dark energy. Higher dimensional theories \cite{highdim} such as
M-theory or string theory may also be helpful to explain this cosmic expansion.
Another justification comes from modification of general theory (GR)
involving some inverse curvature terms \cite{invR}.
However, modified gravity with inverse curvature terms seems
to be unstable and may not pass solar system tests
\cite{unstable}. This discrepancy can be addressed by including higher
derivative terms. Moreover, the viability can be
achieved by considering squared curvature terms \cite{viable}. It has been suggested that the
current expansion may be justified if we add some suitable powers of
curvature in the usual Einstein-Hilbert action \cite{5}.
Thus it seems interesting to study the universe in the
context of alternative or modified theories of gravity.

Among various modification, $f(R)$ gravity
is a possible candidate which gives a natural gravitational
alternative to dark energy \cite{Nojiri Alternative for Dark
Energy}. This theory may provide an easy unification of early time inflation
and late time acceleration. The cosmic acceleration can be explained
by introducing the term $1/R$ at small curvatures.
It was Buchdahl \cite{5B} who introduced $f(R)$ gravity using non-linear Lagrangians.
The $f(R)$ theory of gravity seems most suitable due
to its cosmologically important $f(R)$ models. These models
consist of higher order curvature terms as functions of Ricci
scalar $R$. Some viable $f(R)$ gravity models \cite{5} have been
suggested which show the unification of late-time acceleration and
early-time inflation. It is  now expected that dark matter problem
can be addressed using viable $f(R)$ models. In recent years, many
authors have shown keen interest to investigate this theory in
different context \cite{Analysis of $f(R)$ Theory Corresponding to
NADE and NHDE}-\cite{Noether Symmetry Approach in f(R) Tachyon
Model}. Some detailed reviews are available to better understand the theory \cite{Review Sotiriou}.
Multam$\ddot{a}$ki and Vilja \cite{6} investigated static spherically symmetric vacuum solutions in
$f(R)$ theory. They established that the field equations in $f(R)$
gravity gave the Schwarzschild de Sitter solution. Exact spherically symmetric
interior solutions in metric $f(R)$ gravity have been studied by
Shojai and Shojai \cite{Shojai}. Hollenstein and Lobo \cite{10}
analyzed static spherically symmetric solutions in $f(R)$ gravity
coupled to non-linear electrodynamics. $f(R)$ gravity at one-loop level in de Sitter
universe has been investigated by Cognola et al. \cite{8}.
Cylindrical symmetry has also been widely used to
investigate $f(R)$ gravity in different contexts \cite{Cylindrical thin-shell wormholes in f(R) gravity}.
Recently developed $f(T)$ gravity
is another alternative theory which is the generalization of
teleparallel gravity. This theory also seems interesting as it may explain the cosmic acceleration without
involving the dark energy. A considerable amount of work has been done
in this theory so far \cite{ft}.

Gravitational collapse is an interesting and important issue in GR.
It is involved in the structure formation of the universe causing the existence of
galaxies, stars and planets. The singularity theorem suggests that
the occurrence of spacetime singularity is a
general feature of any cosmological model under some reasonable
conditions. So the solutions with singularities can be
produced by the gravitational collapse of non-singular and asymptotically flat
initial data \cite{G5}. The classification of spacetime singularities is
based on two facts whether they can be observed or not.
If a spacetime singularity is locally observable then it is termed as naked.
A black hole is a spacetime singularity which can not be observed.
Penrose \cite{G8} proposed a cosmic censorship conjecture that the singularities
appearing in the gravitational collapse are always covered by an event horizon.
Formation of compact stellar objects like neutron stars and white dwarf is the result of
gravitational collapse. Spherical symmetry has been extensively used to study
gravitational collapse. Dust gravitational collapse was first explored by
Oppenheimer and Snyder \cite{G9}. Markovic and Shapiro \cite{G15} extended their work
by considering positive cosmological constant. Many researchers \cite{G10}
have investigated gravitational collapse by considering interior and exterior regions.

In the recent years, many authors have shown keen interest to explore
gravitational collapse in alternative theories of gravity \cite{G21}-\cite{G25}.
It has been shown that farther one goes from GR,
there is a greater chance of having a naked singularity \cite{G21}. Ghosh and Maharaj \cite{G26}
obtained a condition for the occurrence of a naked singularity in the collapse of null dust
in higher dimensional $f(R)$ gravity. Openheimer-Snyder collapse in Brans-Dicke theory has
been discussed by Scheel \cite{G026}. In a recent paper \cite{G27}, Rudra and Debnath discussed
gravitational collapse in Vaidya spacetime for Galileon theory of gravity.
Sharif and Abbas \cite{G28} studied the dynamics of shearfree dissipative gravitational collapse in $f(G)$ gravity.
Spherically symmetric perfect fluid gravitational collapse has been discussed in metric $f(R)$ gravity by Sharif and Kausar \cite{G29}.
Cembranos et al. \cite{G30} analyzed a general $f(R)$ model with uniformly collapsing cloud of self-gravitating dust particles.

In this paper, we are focussed to discuss the gravitational
collapse with dust case in $f(R)$ gravity. We take Friedmann-Robertson-Walker (FRW)
spacetime in the interior region and Schwarzschild metric in the exterior region.
The paper is organized as follows: Section $2$ is used to give general formalism about
junction conditions between interior and exterior regions.
We introduce field equations in $f(R)$ gravity and solve them using FRW metric for
dust case in section $3$. In Section $4$, we find the apparent
horizons and discuss the role of constant curvature term. Finally, we summarize the
results in the last section.

\section{General Formalism}

In this section, we give junction conditions at the surface
of a collapsing dust sphere. For this purpose, a 4D spherically symmetric spacetime is divided by a
time-like 3D hypersurface $\Sigma$ into two regions namely
interior and exterior regions. Interior and exterior regions
are denoted by  $V^-$ and $V^+$ respectively. Interior region represented by FRW spacetime is given by
\begin{equation}\label{GC1}
ds^2_-=dt^2-a^2(t)dr^2-a^2(t)b^2(r)[d\theta^2+Sin^2\theta d\phi^2],
\end{equation}
where $a(t)$ is cosmic scale factor and
\[ b(r) = \left\lbrace
  \begin{array}{c l}
    \sin{r}, & \text{~~~~~when~~k=1,}\\
   r, & \text{~~~~~when~~k=0,}\\ \sinh{r}, & \text{~~~~~when~~k=-1.}
  \end{array}
\right. \]
For exterior region $V^+$, we consider the Schwarzschild spacetime
\begin{equation}\label{GC2}
ds^2_+=(1-\frac{2M}{R})dT^2-\frac{1}{1-\frac{2M}{R}}dR^2-R^2[d\theta^2+Sin^2\theta d\phi^2],
\end{equation}
where $M$ is an arbitrary constant. Using Israel junction conditions, we consider that first and
second fundamental forms for interior and exterior spacetimes are same. These conditions are given as:\\\\
$1$. The continuity of first fundamental form over $\Sigma$ provides
\begin{equation}\label{GC3}
(ds^2_+)_\Sigma=(ds^2_-)_\Sigma=(ds^2)_\Sigma,
\end{equation}
\\$2$. The continuity of second fundamental form over $\Sigma$ yields
\begin{equation}\label{GC4}
[K_{ij}]=K^+_{ij}-K^-_{ij}, \quad(i,j=0,2,3),
\end{equation}
where the extrinsic curvature tensor $K_{ij}$ is defined as
\begin{equation}\label{GC5}
K^\pm_{ij}=-n^\pm_\sigma\bigg(\frac{\partial^2
x^\sigma_\pm}{\partial\varepsilon^i\partial\varepsilon^j}+\Gamma^\sigma_{\mu\nu}
\frac{\partial x^\mu_\pm}{\partial\varepsilon^i}\frac{\partial
x^\nu_\pm} {\partial\varepsilon^j}\bigg),\quad(\sigma,\mu,\nu=0,1,2,3).
\end{equation}
Here $\varepsilon^i$ and $x^\sigma_\pm$ correspond to the coordinates on
$\Sigma$ and $V^\pm$ respectively. The christoffel symbols $\Gamma^\sigma_{\mu\nu}$
are calculated using the interior and exterior spacetimes and $n^\pm$
are the components of outward unit normal to $\Sigma$ in the
coordinates $x^\sigma_\pm$. Using interior and exterior spacetimes,
the equations of hypersurface $\Sigma$ are written as
\begin{equation}\label{GC6}
h_-(r,t)=r-r_\Sigma=0,
\end{equation}
\begin{equation}\label{GC7}
h_+(R,T)=R-R_\Sigma(T)=0,
\end{equation}
where $r_\Sigma$ is an arbitrary constant. Using these equations,
interior and exterior  metrics given in Eq.(\ref{GC1})
and(\ref{GC2}) take the form
\begin{equation}\label{GC8}
(ds^2_-)_\Sigma=dt^2-a^2(t)b^2(r_\Sigma)[d\theta^2+Sin^2\theta d\phi^2],
\end{equation}
\begin{equation}\label{GC9}
(ds^2_+)_\Sigma=\bigg[1-\frac{2M}{R_\Sigma}-\frac{1}{1-\frac{2M}{R_\Sigma}}\bigg(\frac{dR_\Sigma}{dT}\bigg)^2\bigg]dT^2
-R_\Sigma^2\bigg[d\theta^2+Sin^2\theta d\phi^2\bigg].
\end{equation}
Here we assume $1-\frac{2M}{R_\Sigma}-\frac{1}{(1-\frac{2M}{R_\Sigma})}(\frac{dR_\Sigma}{dT})^2>0$
so that $T$ remains time-like coordinate. Using junction condition (\ref{GC3}), we get
\begin{equation}\label{GC11}
R_\Sigma=a(t)b(r_\Sigma),
\end{equation}
\begin{equation}\label{GC12}
\bigg[1-\frac{2M}{R_\Sigma}-\frac{1}{1-\frac{2M}{R_\Sigma}}\bigg(\frac{dR_\Sigma}{dT}\bigg)^2\bigg]^{\frac{1}{2}}dT=dt.
\end{equation}
Now using Eqs. (\ref{GC6}) and (\ref{GC7}), the outward unit normals
to the interior and exterior spacetimes are given by
\begin{equation}\label{GC13}
n^-_\mu=(0,a(t),0,0),
\end{equation}
\begin{equation}\label{GC14}
n^+_\mu=(-\dot{R}_\Sigma,\dot{T},0,0),
\end{equation}
while the components of extrinsic curvature $K^\pm_{ij}$ turn out to be
\begin{equation}\label{GC15}
K^-_{00}=0, \quad K^-_{22}=csc^2\theta K^-_{33}=(abb')_\Sigma,
\end{equation}
\begin{equation}\label{GC16}
K^+_{00}=\bigg[\dot{R}\ddot{T}-\dot{T}\ddot{R}+\frac{3M\dot{R}^2\dot{T}}{R(R-2M)}-\frac{M(R-2M)\dot{T}^3}{R^3}\bigg]_\Sigma,
\end{equation}
\begin{equation}\label{GC17}
K^+_{22}=csc^2\theta K^+_{33}=(\dot{T}(R-2M))_\Sigma.
\end{equation}
Here dot and prime denote differentiation with respect to $t$ and $r$ respectively.
By applying continuity condition on extrinsic curvatures, we obtain
\begin{equation}\label{GC18}
K^+_{00}=0,\quad K^+_{22}=K^-_{22}.
\end{equation}
Now using Eqs. (\ref{GC15}-\ref{GC18}) along with (\ref{GC11}) and (\ref{GC12}),
the junction conditions take the form
\begin{equation}\label{GC19}
({\dot{b}}')_{\Sigma}=0, \quad M=\bigg(\frac{ab+a\dot{a}^2b^3-abb'^2}{2}\bigg)_\Sigma
\end{equation}
It is mentioned here that equations (\ref{GC11}), (\ref{GC12})
and (\ref{GC19}) forms necessary and sufficient conditions to match the interior and exterior regions smoothly.

\section{$f(R)$ Gravity and Field Equations}

The metric tensor has a key role in GR. One of the main
features of GR is the dependence of Levi-Civita connection on
the metric tensor. However, the connection does not remain the Levi-Civita connection
if we allow the torsion in the theory.
Consequently the dependence of connection on metric tensor vanishes. This
is the main idea behind different approaches of $f(R)$ theories
of gravity.

We get metric version of $f(R)$ gravity if the connection is
the Levi-Civita connection. In this approach, the variation of
action is done with respect to the metric tensor only. The action for $f(R)$ gravity is
\begin{equation}\label{1}
S_{f(R)}=\int\sqrt{-g}(f(R)+L_{m})d^4x,
\end{equation}
where $f(R)$ is a general function of the Ricci scalar and $L_{m}$
is the matter Lagrangian. It would be worthwhile to mention here
that the standard Einstein-Hilbert action can be achieved when
$f(R)=R$. Varying this action with respect to the metric tensor
yields the modified field equations
\begin{equation}\label{2}
F(R)R_{\mu\nu}-\frac{1}{2}f(R)g_{\mu\nu}-\nabla_{\mu}
\nabla_{\nu}F(R)+g_{\mu\nu}\Box F(R)=\kappa T^m_{\mu\nu}.
\end{equation}
Here $F(R)\equiv df(R)/dR$, $\kappa$ is the coupling constant,
$T^m_{\mu\nu}$ is the standard energy-momentum tensor and
\begin{equation}\label{3}
\quad\Box\equiv\nabla^{\mu}\nabla_{\mu}
\end{equation}
with $\nabla_{\mu}$ is the covariant derivative. We can write the
field equations in an alternative form which is familiar with GR
field equations
\begin{equation}\label{4}
G_{\mu\nu}=R_{\mu\nu}-\frac{1}{2}g_{\mu\nu}R=T^c_{\mu\nu}+\tilde{T}^m_{\mu\nu},
\end{equation}
where $\tilde{T}^m_{\mu\nu}=T^m_{\mu\nu}/F(R)$ and the
energy-momentum tensor for gravitational fluid is
\begin{equation}\label{5}
T^c_{\mu\nu}=\frac{1}{F(R)}\bigg[\frac{1}{2}g_{\mu\nu}\bigg(f(R)-RF(R)\bigg)+
F(R)^{;\alpha\beta}\bigg(g_{\alpha\mu}g_{\beta\nu}-g_{\mu\nu}g_{\alpha\beta}\bigg)\bigg].
\end{equation}
It can be seen from Eq.(\ref{5}) that energy-momentum tensor for
gravitational fluid $T^c_{\mu\nu}$ contributes matter part from
geometric origin. This approach is interesting as it may provide
all the matter components required to investigate the dark
universe. Thus it is hoped that $f(R)$ theory of gravity may give
fruitful results to understand the phenomenon of expansion of
universe. When we contract Eq.(\ref{2}), it follows that
\begin{equation}\label{6}
F(R)R-2f(R)+3\Box F(R)=\kappa T^m,
\end{equation}
where $T^m$ is the trace of energy-momentum tensor.
Here we are interested in pressureless gravitational collapse.
For dust, the energy-momentum tensor is given as
\begin{equation}\label{04}
T^m_{\mu\nu}=\rho u_\mu u_\nu,
\end{equation}
where $\rho$ is the matter density and the four velocity vector
$u_\mu$ satisfies the equation $u_\mu={\delta^0}_\mu$.
Using this equation along with field equations (\ref{2}), we get
three independent differential equations for the interior spacetime
\begin{equation} \label{13}
-\frac{3\ddot{a}}{a}=\frac{1}{F}\bigg[\kappa\rho+\frac{f}{2}-3\frac{\dot{a}\dot{F}}{a}\bigg],
\end{equation}
\begin{equation} \label{14}
\frac{\ddot{a}}{a}+2(\frac{\dot{a}}{a})^2-2\frac{b''}{a^2b}=\frac{1}{F}\bigg[-\frac{f}{2}+2\frac{\dot{a}\dot{F}}{a}+\ddot{F}\bigg],
\end{equation}
\begin{equation} \label{15}
\frac{\ddot{a}}{a}+2(\frac{\dot{a}}{a})^2-\frac{b''}{a^2b}-
(\frac{b'}{ab})^2+\frac{1}{a^2b^2}=\frac{1}{F}\bigg[-\frac{f}{2}+2\frac{\dot{a}\dot{F}}{a}+\ddot{F}\bigg].
\end{equation}
The solution of these highly non-linear differential equations does not seem to be possible
straightforwardly. However, we can try to find a solution using
the assumption of constant scalar curvature, i.e. $R=R_0$. Using this assumption,
left side of Eq.(\ref{6}) becomes constant which leads to a
constant energy density, say $\rho=\rho_0$.
Thus the field equations (\ref{13}-\ref{15}) take the form
\begin{equation} \label{013}
-\frac{3\ddot{a}}{a}=\frac{1}{F(R_0)}\bigg[\kappa\rho_0+\frac{f(R_0)}{2}\bigg],
\end{equation}
\begin{equation} \label{014}
\frac{\ddot{a}}{a}+2(\frac{\dot{a}}{a})^2-2\frac{b''}{a^2b}=-\frac{f(R_0)}{2F(R_0)},
\end{equation}
\begin{equation} \label{015}
\frac{\ddot{a}}{a}+2(\frac{\dot{a}}{a})^2-\frac{b''}{a^2b}-
(\frac{b'}{ab})^2+\frac{1}{a^2b^2}=-\frac{f(R_0)}{2F(R_0)}.
\end{equation}
Manipulating these equations, we obtain
\begin{equation} \label{0014}
2\frac{\ddot{a}}{a}+(\frac{\dot{a}}{a})^2+\frac{1-b'^2}{a^2b^2}=-\frac{1}{2F(R_0)}[\kappa\rho_0+f(R_0)].
\end{equation}
Integrating first equation from (\ref{GC19}), it follows that
\begin{equation} \label{0015}
b'=X,
\end{equation}
where $X$ in an arbitrary integration function of $r$. Thus Eq.(\ref{0014}) takes the form
\begin{equation} \label{00014}
2\frac{\ddot{a}}{a}+(\frac{\dot{a}}{a})^2+\frac{1-X^2}{a^2b^2}=-\frac{1}{2F(R_0)}\bigg[\kappa\rho_0+f(R_0)\bigg].
\end{equation}
Integrating this equation with respect to $t$, we get
\begin{equation} \label{19}
\dot{a}^2=\frac{X^2-1}{b^2}+\frac{2m}{ab^3}-\frac{a^2}{6F(R_0)}\bigg[\kappa\rho_0+f(R_0)\bigg],
\end{equation}
where $m=m(r)$ is an arbitrary function and is related to the mass of the collapsing system
\begin{equation} \label{20}
m(r)=\frac{\kappa\rho_0a^3b^3}{6F(R_0)}.
\end{equation}
Using gravitational units, i.e. $\kappa=8\pi$, the mass of the collapsing system takes the form
\begin{equation} \label{21}
m(r)=\frac{4\pi\rho_0a^3b^3}{3F(R_0)}.
\end{equation}
It is mentioned here that mass of the system must be positive
because negative mass is not acceptable physically.
Using Eq.(\ref{19}) and second junction condition in Eq.(\ref{GC19}), we get
\begin{equation} \label{22}
M=m-\frac{a^3b^3[8\pi\rho_0+f(R_0)]}{12F(R_0)}.
\end{equation}
Now we calculate the total energy $\tilde{M}(r,t)$
at a time $t$ for the interior hypersurface of radius $r$ using the mass function \cite{G030}
\begin{equation} \label{23}
\tilde{M}(r,t)=\frac{ab}{2}[1+g^{\mu\nu}(ab)_{,\mu}(ab)_{,\nu}].
\end{equation}
Using  Eq.(\ref{19}), the mass function turns out to be
\begin{equation} \label{24}
\tilde{M}(r,t)=m(r)-\frac{a^3b^3[8\pi\rho_0+f(R_0)]}{12F(R_0)},
\end{equation}
where $m(r)$ denotes the energy due to constant matter density in Eq. (\ref{21}).
Now we find the solution with $X(r)=1$ using Eq.(\ref{0015}).
In this case the closed form solution turns out to be
\begin{equation} \label{25}
ab=\bigg[\frac{-12mF(R_0)}{8\pi\rho_0+f(R_0)}\bigg]^{\frac{1}{3}}\sinh^{\frac{2}{3}}\alpha(r,t),
\end{equation}
where
\begin{equation} \label{26}
\alpha(r,t)=\sqrt{-\frac{3[8\pi\rho_0+f(R_0)]}{8F(R_0)}}[t_s(r)-t].
\end{equation}
Here we assume $8\pi\rho_0+f(R_0)<0$ to have a realistic solution and
$t_s(r)$ is an arbitary function of $r$. It is clear that $t=t_s$ is the
time formation of singularity for a perticular shell at some distance $r$.
In the limiting case when $f(R_0)\rightarrow -8\pi\rho_0$, the above solution takes the form
\begin{equation} \label{27}
\lim_{f(R_0)\rightarrow -8\pi\rho_0}ab=\bigg[\frac{9m}{2}(t_s-t)^2\bigg]^{\frac{1}{3}},
\end{equation}
which correspond to the well known Tolman-Bondi solution  \cite{G31}.

\section{Apparent Horizons}

We obtain the apparent horizon when the boundary of two trapped spheres is
formed. In this section, we find such boundary of two trapped  spheres whose
outward normals are null. For the interior spacetime (\ref{GC1}), this is given as
\begin{equation} \label{28}
g^{\mu\nu}(ab)_{,\mu}(ab)_{,\nu}=\dot{a}^2b-b'^2=0.
\end{equation}
Using Eq.(\ref{19}) in this equation, we get
\begin{equation} \label{29}
\frac{1}{F(R_0)}[8\pi\rho_0+f(R_0))]a^3b^3+6ab-12m=0.
\end{equation}
The solutions of this equation for $ab$ yield the
apparent horizons. For $f(R_0))=-8\pi\rho_0$, it becomes the
Schwarzschild horizon, i.e., $ab = 2m$. When $m =0$, it
yields a de-Sitter horizon, i.e.
\begin{equation}\label{30}
ab=\sqrt{\frac{-6F(R_0)}{8\pi\rho_0+f(R_0)}}
\end{equation}
The case $3m<\sqrt{\frac{-2F(R_0)}{8\pi\rho_0+f(R_0)}}$
leads to two horizons,
\begin{equation}\label{31}
(ab)_c=\sqrt{\frac{-8F(R_0)}{8\pi\rho_0+f(R_0)}}\cos \frac{\psi}{3}
\end{equation}
and
\begin{equation}\label{32}
(ab)_{bh}=-\sqrt{\frac{-8F(R_0)}{8\pi\rho_0+f(R_0)}}\bigg[\cos \frac{\psi}{3}-\sqrt{3}\sin \frac{\psi}{3}\bigg],
\end{equation}
where the subscripts  $c$ and $bh$ represent cosmological and black hole horizons respectively and
\begin{equation}
\cos \psi=-3m\sqrt{\frac{-2F(R_0)}{8\pi\rho_0+f(R_0)}} .
\end{equation}
If we take $m=0$, the equations (\ref{31}) and (\ref{32}) reduce to
\begin{equation}
(ab)_c=\sqrt{\frac{-6F(R_0)}{8\pi\rho_0+f(R_0)}}, \quad (ab)_{bh}=0.
\end{equation}
It is mentioned here that the results can be generalized when $m\neq 0$ and $8\pi\rho_0+f(R_0)\neq0$ \cite{G32}.\\\\
For the case when $3m=\sqrt{\frac{-2F(R_0)}{8\pi\rho_0+f(R_0)}}$, both horizons coincide, i.e.,
\begin{equation}
(ab)_c=(ab)_{bh}=\sqrt{\frac{-2F(R_0)}{8\pi\rho_0+f(R_0)}}.
\end{equation}
Thus the range of the cosmological
horizon and the black hole horizon becomes
\begin{equation}\label{34}
0\leq(ab)_{bh}\leq\sqrt{\frac{-2F(R_0)}{8\pi\rho_0+f(R_0)}}\leq(ab)_c\leq\sqrt{\frac{-6F(R_0)}{8\pi\rho_0+f(R_0)}}.
\end{equation}
There does not exist any apparent horizon in the case  $3m>\sqrt{\frac{-2F(R_0)}{8\pi\rho_0+f(R_0)}}$.\\\\
The formation time of the apparent horizon can be calculated using Eqs. (\ref{25}) and (\ref{29}) and is
given by
\begin{equation}\label{35}
t_n=t_s-\sqrt{\frac{-8F(R_0)}{3(8\pi\rho_0+f(R_0))}}\sinh^{-1}\bigg[\frac{(ab)_n}{2m}-1\bigg]^\frac{1}{2},\quad n=1,2.
\end{equation}
In the limiting case when $f(R_0)\rightarrow -8\pi\rho_0$, the result corresponds to Tolman-Bondi solution
\begin{equation}
t_{ah}=t_s-\frac{4}{3}m.
\end{equation}
From Eq.(\ref{35}), it follows that
\begin{equation}
\frac{(ab)_{n}}{2m}=\cosh ^2\alpha _n,
\end{equation}
where
\begin{equation}
\alpha _n(r,t)=\sqrt{\frac{-3(8\pi\rho_0+f(R_0))}{8F(R_0)}}[t_s(r)-tn].
\end{equation}
It is clear from Eq.(\ref{35}) that both the
black hole horizon and the cosmological horizon
form earlier than the singularity $t=t_s$.
Now since Eq. (\ref{35}) yields $t_1\leq t_2$ and
Eq. (\ref{34}) implies that $(ab)_{bh}\leq(ab)_c$. This is an indication
that cosmological horizon forms earlier than the black hole horizon.

\section{Concluding Remarks}

This paper is devoted to discuss gravitational collapse in $f(R)$ gravity.
For this purpose, we consider the metric approach of
this theory to study the field equations.
It is observed that the field equations (\ref{13})-(\ref{15})
are complicated and highly non-linear. Thus it seems
difficult to solve them analytically without any assumption.
The assumption of constant curvature has been used to find a solution.

We investigate the gravitational collapse of dust by considering FRW
spacetime as the interior region while for the
exterior region we take Schwarzschild metric. The junction
conditions have been derived between interior and exterior spacetimes.
We get two physical apparent horizons namely cosmological horizon and black hole horizon.
It is found that formation time for black hole horizon is more as compared to cosmological
horizon. Moreover, both horizons are formed earlier than singularity.
This indicates that the singularity is covered, i.e., black hole, which shows that $f(R)$ gravity
supports cosmic censorship conjecture. From dynamical equation (\ref{19}), the rate of gravitational collapse is
\begin{equation}\label{100}
\ddot{a}b=-\frac{m}{a^2b^2}-\frac{ab}{6F(R_0)}[8\pi\rho_0+f(R_0)].
\end{equation}
The acceleration should be negative for the collapsing process which is possible when
\begin{equation}
ab<\bigg[-\frac{6mF(R_0)}{8\pi\rho_0+f(R_0)}\bigg]^{\frac{1}{3}}.
\end{equation}
Thus Eq. (\ref{100}) indicates that $f(R_0)$ slows down the collapsing process when $f(R_0)+8\pi\rho_0<0$.
Further, the presence of $f(R_0)$ causes two physical horizons.
One is the black hole horizon and the other is cosmological horizon.
It also influences the time difference
between the formation of the apparent horizon and
singularity. It is concluded that $f(R_0)$
affects the process of collapse and hence it limits the
size of the black hole. It would be worthwhile to mention hare that the term
$f(R_0)$ plays the same role as that of the cosmological constant in
GR field equations and our results agree with \cite{G10Z}.\\\\
\vspace{1.0cm}
\\\\\textbf{Acknowledgement}\\\\ MFS is thankful to National University
of Computer and Emerging Sciences (NUCES) Lahore Campus, for
funding the PhD programme. The authors are also grateful to the anonymous reviewer
for valuable comments and suggestions to improve the paper

\end{document}